\documentclass[referee, sn-apa]{sn-jnl}

\usepackage{graphicx}%
\usepackage{multirow}%
\usepackage{amsmath,amssymb,amsfonts}%
\usepackage{amsthm}%
\usepackage{mathrsfs}%
\usepackage[title]{appendix}%
\usepackage{xcolor}%
\usepackage{textcomp}%
\usepackage{manyfoot}%
\usepackage{booktabs}%
\usepackage{algorithm}%
\usepackage{algorithmicx}%
\usepackage{algpseudocode}%
\usepackage{listings}%
\usepackage[small]{titlesec}

\renewcommand*{\d}{\mathop{}\!\mathrm{d}}

\begin{document}

\title[Diffeomorphism Invariance]{Diffeomorphism Invariance and General Relativity}

\author*{\fnm{Max} \sur{Heitmann}}

\affil{\orgdiv{Faculty of Philosophy}, \orgname{University of Oxford}\footnote{This is the institution at which most of the research for the present work was conducted. The author is temporarily unaffiliated. Correspondence to: maxfheitmann@gmail.com}}


\abstract{Diffeomorphism invariance is often considered to be a hallmark of the theory of general relativity (GR). But closer analysis reveals that this cannot be what makes GR distinctive. The concept of diffeomorphism invariance can be defined in two ways: under the first definition (diff-invariance\textsubscript{1}), \emph{both} GR and all other classical spacetime theories turn out to be diffeomorphism invariant, while under the second (diff-invariance\textsubscript{2}), \emph{neither} do. Confusion about the matter can be traced to two sources. First, GR is sometimes erroneously thought to embody a ``general principle of relativity,'' which asserts the relativity of all states of motion, and from which it would follow that GR must be diff-invariant\textsubscript{2}. But GR embodies no such principle, and is easily seen to violate diff-invariance\textsubscript{2}. Second, GR is unique among spacetime theories in requiring a general-covariant formulation, whereas other classical spacetime theories are typically formulated with respect to a preferred class of global coordinate systems in which their dynamical equations simplify. This makes GR's diffeomorphism invariance (in the sense of diff-invariance\textsubscript{1}) manifest, while in other spacetime theories it lies latent---at least in their familiar formulations. I trace this difference back to the fact that the spacetime structure is inhomogeneous within the models of GR, and mutable across its models. I offer a formal criterion for when a spacetime theory possesses immutable spacetime structure, and using this criterion I prove that a theory possesses a preferred class of coordinate systems applicable across its models if and only if it possesses immutable spacetime structure.}

\keywords{spacetime, general relativity, diffeomorphism invariance, background independence}

\maketitle

\section{Introduction and Preliminaries}

Diffeomorphism invariance, understood as invariance of the solution
space under active smooth transformations of the fields on spacetime, is
properly understood as a property of a spacetime \emph{theory}, not of
its formulation. In general, formulations stand in a many-to-one
correspondence with theories, and so the properties of a particular
theory formulation may not always be meaningfully predicated of the
theory as a whole. Bearing this in mind, I propose two definitions:
according to the first (diff-invariance\textsubscript{1}), both general
relativity (GR) and all other classical spacetime theories are
diffeomorphism invariant; according to the second
(diff-invariance\textsubscript{2}), neither are. In either case,
diffeomorphism invariance \emph{per se} cannot be what is distinctive
about GR. Rather, I argue that GR's unique association with
diffeomorphism invariance arises from the fact that it \emph{must} be
formulated using general-covariant equations, whereas other spacetime
theories need not be. This general-covariant formulation makes GR's
diff-invariance\textsubscript{1} manifest, while in other spacetime
theories it lies latent (in their familiar formulations). The reason for
this difference is that GR lacks, where other spacetime theories
possess, a preferred class of global coordinate systems to which the
equations can be referred and in which they simplify. Moreover, GR has
this distinctive feature not because it embodies a ``general principle
of relativity'' that asserts the equivalence of all states of motion,
but rather because in GR the spacetime structure is mutable across
models and (generically) highly inhomogeneous within models. Thus, I
shall argue, it is the mutability and inhomogeneity of the postulated
spacetime that constitutes the truly distinctive feature of GR.

First, a little bit of set-up. All the spacetime theories that we will
consider have kinematically possible models (KPMs) of the form
\(\mathfrak{M =}\left\langle M,O_{1},\ldots,O_{n} \right\rangle\), where
\(M\) is a four-dimensional manifold and the \(O_{i}\) are geometrical
object fields defined on \(M\) (i.e., functions, vector fields,
co-vector fields, and tensor fields). Every spacetime theory has a set
of field equations which pick out a subset of the KPMs as representing
physically possible histories; the elements of this subset are the
dynamically possible models (DPMs). In general, the set of DPMs comes
partitioned into equivalence classes of isomorphic models, and it is the
elements of this partition that stand in one-to-one correspondence with
physically possible histories of events in spacetime. I propose to
individuate theories according to what they deem physically
possible---that is, by their associated sets of physically possible
histories. This gives rise to a many-to-one correspondence between
formulations and theories, as two formulations of the same theory may
differ in which equivalence class of DPMs they use to represent each
physically possible history.

Diffeomorphisms of the manifold are denoted \(h\ :M \rightarrow M\). We
will denote the drag-along of the geometrical object \(O_{i}\) under
\(h\) as \(h*O_{i}\), which is nicely ambiguous between the pullback
\(h_{*}\) (appropriate for functions, co-vector fields, and covariant
tensor fields), the pushforward \(h^{*}\) (appropriate for vector fields
and contravariant tensor fields), and the mixed mapping that can be
defined for mixed tensor fields. For any diffeomorphism and any
spacetime theory \(T\), we define the ``lift'' of the diffeomorphism
(denoted \(h*\mathfrak{M}\)) as the transformation effected on the KPMs
by dragging-along some or all of the geometric object fields postulated
by \(T\) (more on this later). \(T\) is then called diffeomorphism
invariant if this lifted transformation preserves the set of DPMs.

\section{Diffeomorphism invariance is not what is distinctive of GR}
Before addressing the question of what really does distinguish GR from
other classical spacetime theories, we ought to get clear on why GR's
diffeomorphism invariance is not up to the task. After all, many authors
seem to state or imply that diffeomorphism invariance is indeed the
property that points to GR's distinctive character.\footnote{For
  example, \cite{Rovelli2007} writes: `Diffeomorphism invariance is the key
  property of the mathematical language used to express the key
  conceptual shift introduced with GR.'} But, of course, if \emph{any} classical spacetime
theory can be given a diffeomorphism-invariant formulation, then this
cannot be the case. Why should we grant that any classical spacetime
theory can be given a diffeomorphism-invariant formulation? Consider the
following argument, which I will call the ``Argument from General
Covariance.'' \\

\emph{Argument from General Covariance}:

\begin{enumerate}
\def\labelenumi{\arabic{enumi}.}
\item
  Any classical spacetime theory can be given a coordinate-invariant or
  ``generally covariant'' formulation. This means that the equations are
  written in such a way that they pick out exactly the same class of
  DPMs in any coordinate system.
\item
  Corresponding to any diffeomorphism $ h : M \rightarrow M $ there
  exists a coordinate transformation
  \(\left\langle x_{i} \right\rangle \rightarrow \left\langle y_{i} \right\rangle\)
  (given by \(y_{i} = x_{i} \circ h\)) such that the components of
  the geometric object fields with respect to
  \(\left\langle y_{i} \right\rangle\) in \(h*\mathfrak{M}\) are the
  same as the components of the corresponding object fields with respect
  to \(\left\langle x_{i} \right\rangle\) in \(\mathfrak{M}\). Thus, if
  the fields in \(\mathfrak{M}\) satisfy the equations of the theory
  when referred to the coordinates \(\left\langle x_{i} \right\rangle\),
  the fields in \(h*\mathfrak{M}\) will satisfy the equations when
  referred to \(\left\langle y_{i} \right\rangle\).
\item
  But since we can formulate our theory in a general-covariant way, the
  equations pick out the same class of DPMs when referred to either
  \(\left\langle x_{i} \right\rangle\) or
  \(\left\langle y_{i} \right\rangle\). And so, in particular, if
  \(h*\mathfrak{M}\) is a DPM according
  \(\left\langle y_{i} \right\rangle\) then it is also a DPM according
  to \(\left\langle x_{i} \right\rangle\), and indeed according to every
  other coordinate system. Thus, the general-covariant formulation
  reveals that \(h*\mathfrak{M}\) is a model of the theory if
  \(\mathfrak{M}\) is---in other words, the theory is diffeomorphism
  invariant.
\end{enumerate}

In essence, this argument is simply pointing out that diffeomorphisms
can be interpreted as ``active'' coordinate transformations, and whether
a coordinate transformation is active or passive cannot make a
\emph{formal} difference: if a passive transformation leaves the
equations of motion satisfied, then so will an active transformation.
Premise (1) asserts that any classical spacetime theory can be
formulated in such a way that passive transformations leave the
equations of motion satisfied. Thus, since it is \emph{possible} to
formulate all classical spacetime theories in this way, they must all in
fact be diffeomorphism invariant.

What should we make of this argument? We might first look to reject
premise (1), which appears to be doing most of the work. But this
premise is supported by several compelling considerations. First, there
is the theoretical consideration that coordinates are merely labels for
physical events and as such do not carry any direct physical
significance, whereas plausibly only laws that relate the physically
significant elements of the theory can play a role in determining which
histories are physically possible. Thus, since the set of DPMs
represents the set of histories that the theory deems physically
possible, we should expect the theory to be able to circumscribe that
set in a coordinate-invariant manner. Second, there is the practical
consideration that every classical spacetime theory that has ever been
devised \emph{has been} given a general-covariant formulation. In
practice, the way one does this is to introduce onto the manifold
geometrical object fields corresponding to all the elements of spacetime
structure that are needed to ground the absolute distinctions between
different states of motion implied by the laws of the theory; the
equations of motion can then be formulated with reference to these
objects without assuming that the coordinate system being used is
``well-adapted'' to them, or indeed (if preferred) without using any
coordinate system at all.\footnote{These remarks echo the well-known ``Kretschmann objection''; for discussion, see \cite{Pooley2015}.}

For example, it is well-known that the requisite spacetime structure
needed to ground the laws of Newtonian kinematics is not Newton's
absolute space and time but rather \emph{Galilean spacetime}: this
spacetime is sufficiently well structured to distinguish absolutely
between inertial and accelerated motion (as required by Newton's first
two laws), but not so highly structured as to distinguish between
different states of inertial motion (which do not make a dynamical
difference). To represent this mathematically, we therefore introduce
onto our manifold an affine connection \(D\) but \emph{not} a
``rigging'' of the manifold with a preferred congruence of time-like
geodesics. Models of this spacetime theory take the form
\(\left\langle M,D, \d t,h,\sigma \right\rangle\), where \(D\) is the
affine connection, \( \d t\) is a co-vector field giving the temporal
metric, \(h\) is a degenerate symmetric tensor field giving the spatial
metric on the simultaneity hyperplanes, and \(\sigma\) is a class of
curves representing the trajectories of particles through the spacetime.
We further introduce field equations to constrain these geometrical
objects in various ways, for example by requiring the connection \(D\)
to be flat and metric compatible. We are then in a position to write
Newton's first law, which says that free particles follow time-like
geodesics, in the coordinate free form \(D_{T_{\sigma}}T_{\sigma} = 0\),
where \(T_{\sigma}\) is the tangent vector field to the worldline
\(\sigma\) of the free particle \citep[pp. 71--92]{Friedman1983}. Thus, both for reasons of principle and because it has been
demonstrated in practice, we are obliged to accept premise (1).

The more promising place to look for problems is premise (2). In
particular, premise (2) makes a hidden assumption which plays a role in
the definition of the ``lift'' of the diffeomorphism presupposed in the
argument. To appreciate this, notice that the geometric object fields
appearing in the KPMs of a spacetime theory come divided into two
classes: those fields (such as the metric field) which represent
spacetime structure, and those fields which represent the material
contents of spacetime. To highlight this, we can write a generic KPM as
\(\left\langle M,S,P \right\rangle\) where \(S\) is a sequence of tensor
fields representing the spacetime structure and the \(P\) is a sequence
of tensor fields representing the configuration of matter in spacetime.
Given a KPM \(\mathfrak{M =}\left\langle M,S,P \right\rangle\) and a
diffeomorphism \( h : M \rightarrow M \), premise (2) implicitly assumed
that the lift \(h*\mathfrak{M}\) ought to be defined as
\(\left\langle M,h*S,\ h*P \right\rangle\). But if we are to do justice
to the idea of diffeomorphisms as genuinely \emph{active}
transformations which shuffle the fields around spacetime, such a
definition is only appropriate if spacetime is correctly represented by
the bare manifold \(M\). Alternatively, we might think that spacetime is
better represented by the tuple
\(\left\langle M,S \right\rangle\)---i.e., by the manifold together with
the structure fields. In this case, an active diffeomorphism would be
represented mathematically by keeping the structure fields pinned down
to the manifold rather than by dragging them along with the matter
fields; we would define the lift as
\(h*\mathfrak{M}\ \  = \left\langle M,S,h*P \right\rangle\) accordingly.

We do not need to adjudicate the dispute over whether it is \(M\) or
\(\left\langle M,S \right\rangle\) that should be taken to represent
spacetime in our models. For our purposes, it suffices to note that
these two possibilities give rise to two different notions of
diffeomorphism invariance. According to the first, which I will call
diff-invariance\textsubscript{1}, a theory \(T\) is diffeomorphism
invariant iff, if \(\left\langle M,S,P \right\rangle\) is a model of the
theory, then so is \(\left\langle M,h*S,h*P \right\rangle\) for all
diffeomorphisms \(h\). According to the second, which I will call
diff-invariance\textsubscript{2}, a theory \(T\) is diffeomorphism
invariant iff, if \(\left\langle M,S,P \right\rangle\) is a model of the
theory, then so is \(\left\langle M,S,h*P \right\rangle\) for all
diffeomorphisms \(h\). The argument from general covariance establishes
that any theory which can be given a general-covariant formulation will
be diff-invariant\textsubscript{1}. Since both GR and all other
classical spacetime theories can be given a general-covariant
formulation, both must be diff-invariant\textsubscript{1}, and so this
notion of diffeomorphism invariance does not characterise what is
distinctive about general relativity.

On the other hand, it should be obvious that many of the major pre-GR
classical spacetime theories \emph{fail} to be
diff-invariant\textsubscript{2}. For example, Newtonian mechanics and
special relativity both incorporate affine structure, requiring that
freely moving particles follow affine geodesics. Suppose we have a very
simple model of such a theory, consisting of an inertially moving
particle in an otherwise empty spacetime. A lifted diffeomorphism that
passes over the affine structure field and acts only on the matter
fields would allow us to smoothly deform the worldline of that particle
into a curve \emph{without simultaneously changing the standard of
straightness}. The resulting KPM obviously does not represent a physical
possibility---it is not a DPM. But general relativity \emph{also} fails
to be diff-invariant\textsubscript{2}. Since in general relativity all
the spacetime structure is derived from the metric tensor \(g\), the
kinematically possible models take the form
\(\left\langle M,g,\Phi \right\rangle\), where \(g\) is a structure
field and \(\Phi\) represents any matter fields.
Diff-invariance\textsubscript{2} therefore requires that, if
\(\left\langle M,g,\Phi \right\rangle\) is a dynamically possible model,
then so is \(\left\langle M,g,h*\Phi \right\rangle\). But this
requirement is obviously not satisfied by general relativity: such a
lifted diffeomorphism generically breaks the connection between the
metric and matter fields in both directions, leading to violations of
both the Einstein field equations and the geodesic equation.

In sum, we have seen that if we view the structure fields as dragged
along with the matter fields under diffeomorphisms, then both GR and all
other classical spacetime theories are diffeomorphism invariant
(diff-invariance\textsubscript{1}), whereas if we view the structure
fields as pinned down to the manifold and only drag along the matter
fields, then neither GR nor several major pre-GR spacetime theories are
diffeomorphism invariant (diff-invariance\textsubscript{2}). At this
stage it might be urged that there is a third option
(diff-invariance\textsubscript{3}) which does manage to drive a wedge
between GR and its predecessors: view the structure fields as pinned
down to the manifold just when they are non-dynamical, and as dragged
along with the matter fields just in case they are dynamical (Rovelli, \citeyear{Rovelli2001}, as quoted by Pooley, \citeyear{Pooley2015}, appears to advocate this strategy).
Then, since GR is the only spacetime theory to have explicitly dynamical
spacetime structure,\footnote{At least, the only \emph{traditionally
  formulated} spacetime theory to have dynamical spacetime structure.
  The equivalence of gravitational and inertial mass actually allows for
  a reformulation of Newtonian gravitation theory which does away with
  ``gravitational force'' in favour of a dynamically modified affine
  connection \citep[pp. 95--104]{Friedman1983}.} we would apply the first
definition to GR only and conclude that GR (and GR \emph{alone}) is
diff-invariant\textsubscript{3}.

Unfortunately, while initially promising, we find that the verdicts
delivered by this third definition are not so clear-cut after all. The
problem is that we can always ``unfix'' non-dynamical structure fields
by no longer giving them outright as prior geometry, but instead
introducing them as dynamical fields and adding field equations to
constrain them to the desired values. For example, in the standard
formulation of special relativity, the kinematically possible models are
of the form \(\left\langle M, \eta, \Phi \right\rangle\), where \(M\)
and \(\eta\) are given outright and the variation between KPMs involves
only variation of the matter fields \(\Phi\) defined on this
``background'' structure. In this formulation \(\eta\) is non-dynamical,
and so the definition under consideration (diff-invariance\textsubscript{3}) would deliver the verdict that
special relativity is not diffeomorphism invariant, because it is not
diff-invariant\textsubscript{2}. On the other hand, we can reformulate
special relativity so that the KPMs are of the form
\(\left\langle M,g,\Phi \right\rangle\)---the \emph{same} as the KPMs of
general relativity---and add the equation \(R_{\ \ bcd}^{a}(g) = 0\) to
constrain the metric to be flat. To every DPM of the standard
formulation corresponds a class of diffeomorphic DPMs in this
reformulated version \citep{Pooley2015}. But since all the models in a given
class of diffeomorphic DPMs are isomorphic to each another, this can
plausibly be interpreted as representational redundancy; plausibly, the
reformulated version of special relativity picks out the same class of
\emph{physically possible histories} as the standard version, and so
under our definition counts as the same theory. But in this version
\(g\) is dynamical---at least insofar as it is determined by a field
equation rather than given outright in the KPMs---and so the definition
under consideration would deliver the verdict that special relativity
\emph{is} diffeomorphism invariant, because it is
diff-invariant\textsubscript{1}.

Since our interest is in what is distinctive about general relativity
\emph{as a theory}, I take it as a severe disadvantage of this third
definition that the verdicts it delivers can vary under mere
reformulations of the spacetime theory in question. Such a definition
could at most tell us what is distinctive about the way GR is
\emph{formulated}, rather than telling us what is distinctive about GR
itself. By contrast, the two definitions
diff-invariance\textsubscript{1} and diff-invariance\textsubscript{2}
are not sensitive to formulation: holding fixed the choice among the two
definitions, a theory cannot be made diff-invariant (or made
not-diff-invariant) via a mere reformulation. Thus,
diff-invariance\textsubscript{1,2} express genuine properties of
spacetime theories. But neither of them can be used to drive a wedge
between GR and other classical spacetime theories, and so neither is up
to the task of characterising what is distinctive about GR.

\section{What is distinctive about general relativity?}
If diffeomorphism invariance \emph{per se} does not set GR apart from
other classical spacetime theories, why is it so frequently considered
to be the mathematical expression of GR's distinctive character? In
brief, I think association runs as follows. GR's formulation in terms of
general-covariant equations makes it manifestly clear that it is
diff-invariant\textsubscript{1} (via the considerations presented in the
Argument from General Covariance). On the other hand, even though other
classical spacetime theories \emph{can} also be formulated using
general-covariant equations, the fact is that they are not standardly
formulated in this way; thus, although these other spacetime theories
are also diff-invariant\textsubscript{1}, this fact is not
\emph{manifest} in their standard mathematical formulations. The reason
for this difference in the standard formulation of GR as compared to
other classical spacetime theories is that in formulating the latter we
can avail ourselves of the convenience of a simplified formalism by
referring the equations to a certain class of preferred coordinate
systems, whereas in GR we are \emph{forced} to formulate the equations
in a general-covariant way due to a lack of preferred coordinate
systems. It is therefore this lack of preferred coordinate systems that
is the truly distinctive feature of GR.

\subsection{Against the ``general principle of relativity''}
At this stage, however, we must ward off a serious misunderstanding that
can arise in connection with the claim that GR ``lacks preferred
coordinates.'' The freedom to choose different coordinate systems is
often associated with a relativity principle, which asserts the
relativity or non-absoluteness of different states of motion. For
example, the freedom to choose any inertial frame of reference in
Newtonian physics is associated with the Galilean principle of
relativity, which asserts that all uniform straight-line motion is
relative motion: there are no absolute standards of rest and no absolute
facts about which objects are ``really'' moving and how fast they are
moving. Thus, we can equally well attach a frame of reference to any
inertially moving object, and the Newtonian physics as described in any
of the resulting coordinate systems should be the same. The same is true
of the freedom to choose any inertial frame in special relativity. Thus,
the relativity of certain states of motion in Newtonian physics and
special relativity is bound up with the freedom to choose any from among
a preferred class of coordinate systems to describe the physics of those
theories.

To see the connection more clearly, consider the passage between ever
more lightly structured classical spacetimes. Recall that full Newtonian
spacetime postulates KPMs of the form
\(\left\langle M,D,\d t,h,V,\sigma \right\rangle\). As we move from full
Newtonian spacetime to neo-Newtonian or Galilean spacetime, we remove
the ``rigging'' of the spacetime, represented in the KPMs by a
covariantly constant time-like vector field \(V\). Thus, it no longer
becomes meaningful to ask about an object's state of absolute rest or
motion---all inertial motion becomes \emph{relativised} to other
inertially moving bodies. At the same time, it is no longer possible to
``adapt'' our coordinate systems so that the coordinate curves of
constant spatial position lie along \(V\), and so the class of preferred
coordinate systems expands to include all those whose coordinate curves
of constant spatial position lie along the time-like geodesics of the
affine connection \(D\)---i.e., all the inertial frames. Next, as we
move from Galilean spacetime to Leibnizian spacetime, we further remove
the inertial structure from our spacetime, represented in the KPMs by
the connection \(D\). Now, it is no longer even meaningful to ask about
an object's state of absolute acceleration; the only meaning that can be
given to ``accelerated motion'' is acceleration \emph{relative} to some
other bodies. And again, the removal of the inertial structure means
that it is no longer possible to adapt our coordinate systems to \(D\),
and so the class of preferred coordinate systems is forced to expand
further to include all \emph{rigid} frames---i.e., those which are
adapted to the spatial metric tensor \(h\), for which constant
coordinate distance over time corresponds to constant spatial distance
over time. As we remove structure from a spacetime, more and more states
of motion become relativized or non-absolute, and the associated class
of ``well-adapted'' coordinate systems expands in lockstep \citep[for a clear presentation, see][]{Earman1989}.

A natural thought would then be that the freedom to choose
\emph{absolutely any} coordinate system in GR comes associated with a
\emph{general} principle of relativity (GPR) asserting the relativity of
\emph{absolutely all} states of motion. If GPR were to obtain in GR,
then it would follow that the set of spacetime structure fields \(S\) is
empty in the theory. For whenever the set of spacetime structure fields
is not empty, these fields can be used to define distinctions between
different absolute states of motion (as above), in contradiction to
GPR's assertion that all distinctions between different states of motion
are distinctions between different states of relative motion. But if the
set of spacetime structure fields is empty, then
diff-invariance\textsubscript{2} collapses into
diff-invariance\textsubscript{1}, and so any theory in which GPR obtains
is not only diff-invariant\textsubscript{1} (as all our theories are),
but in fact diff-invariant\textsubscript{2} as well! Thus, if GPR were
to obtain in GR, we would get a delightfully simple statement of what
makes GR distinctive---namely, that it is the only spacetime theory ever
seriously proposed that is diff-invariant\textsubscript{2}.

But this is all completely wrong. General relativistic spacetime is
actually quite highly structured, sufficient to support a non-relational
distinction between, for example, geodesic and non-geodesic motion (in
violation of the GPR), and to set up preferred ``locally inertial''
coordinate systems in small patches of spacetime. General relativistic
spacetime is, in fact, just as highly structured as the Minkowski
spacetime of special relativity in terms of the spacetime structure
fields it contains (both explicitly and derived)---namely, a metric
field, and a compatible affine connection. So none of the ideas in the
previous paragraph are even remotely on target: GR does not embody GPR,
the set of spacetime structure fields is not empty in the theory, and unfortunately
diffeomorphism invariance really isn't of any help in articulating GR's
distinctive features.

\subsection{The inhomogeneity and mutability of spacetime}
In what sense, then, does GR ``lack preferred coordinates''? GR lacks a
non-trivial class of preferred coordinate systems not because it
embodies GPR and consequently because all coordinate systems are equally
preferred, but rather quite the opposite: in general, all coordinate
systems are equally \emph{dis}-preferred. General relativistic spacetime
has plenty of structure to which we'd want to adapt our
coordinate systems, but in virtue of the inhomogeneity of that structure
there is generically no consistent way of doing this. Consider again
inertial structure. In every local patch of spacetime we can set up
``adapted'' coordinates in which linear equations in the coordinates
represent affine geodesics of the connection. But typically, we cannot
extend this coordinate system to cover the whole spacetime while
retaining this desirable property: a generic model of general
relativistic spacetime lacks global geodesic congruences that could be
used to define the coordinate curves of a preferred coordinate system
adapted to the affine structure. Thus, the class of preferred global
coordinate systems is trivial in GR not because it includes all possible
coordinate systems, but because it includes none.

To avoid confusion, we should distinguish two statements that can be
made about this. The first is that a generic model of GR lacks a
preferred class of coordinate systems (i.e., ones that are globally
adapted to the spacetime structure). The second is that GR \emph{as a
theory} lacks a preferred class of coordinate systems. Both statements
are true about GR, but for different reasons: a generic model lacks a
preferred class of coordinate systems because, as explained above, it is
sufficiently \emph{inhomogeneous} so as to lack (for example) a global
geodesic congruence of curves. But some particularly tidy models of GR
(e.g., vacuum GR, the Schwarzschild solution) can clearly be equipped
with a preferred class of coordinate systems. Nevertheless, GR \emph{as
a theory} still lacks one because the spacetime structure postulated by
GR is \emph{mutable} and so coordinates which are well-adapted to the
spacetime structure of one model will not necessarily be well-adapted to
a different model. Of course, the first statement entails the second,
but importantly the second statement would still hold true even if
every model of GR could be individually equipped with globally
well-adapted coordinates, due to the fact that the spacetime structure
is mutable in GR and so no single coordinate system will be well-adapted
\emph{across} the models.

It would be nice to have a formal criterion for this ``mutability''
across models. To this end, I will appropriate Earman's definition of
\emph{similarity}: for a subset \(\Theta\) of the geometric object types
postulated by a theory \(T\), we say that \(\Theta\) \emph{remains
similar} for \(T\) just in case for any models
\(\left\langle M,O_{1},\ldots,O_{n} \right\rangle\),
\(\left\langle M,O_{1}^{'},\ldots,O_{n}^{'} \right\rangle\) of \(T\)
there exists a diffeomorphism \(h : M \rightarrow M\) such that
\(h*O_{i} = O_{i}^{'}\) for each of the geometric object types
\(\mathbf{O}_{\mathbf{i}} \in \Theta\) \citep[p. 38]{Earman1989}.\footnote{For
  the avoidance of confusion: boldface \(\mathbf{O}_{\mathbf{i}}\)
  represents an object \emph{type,} tokenings of which are present in
  every KPM and are represented by non-boldface \(O_{i},O_{i}^{'}\).} I
will then say that a theory \(T\) possesses \emph{immutable spacetime
structure} just in case, for some non-empty
\(\Theta \subset \mathbf{S}\), \(\Theta\) remains similar for \(T\). The
motivation behind this definition is the desire to capture a sense of
the ``sameness'' of a structure field across DPMs in recognition of the
fact that, as discovered earlier, all spacetime theories are at least
diff-invariant\textsubscript{1}.

This formal idea of immutable spacetime structure is well-suited to the
task at hand. For suppose a theory \(T\) possesses immutable spacetime
structure. Then \(T\) postulates a set of object types
\(\Theta \subset \mathbf{S}\) which remains similar for \(T\). Suppose
these object types appear in a given model \(\mathfrak{M}\) of \(T\) as
the geometric object fields \(\text{\{}O_{i}\text{\}}\), and suppose
that in this model there exists a coordinate system (or class of
coordinate systems) which is globally well-adapted to
\(\text{\{}O_{i}\text{\}}\). Then also in any other model
\(\mathfrak{N}\) of \(T\) we are guaranteed to find a ``counterpart''
coordinate system (or class of coordinate systems) which is globally
well-adapted to the corresponding set \(\text{\{}O_{i}^{'}\text{\}}\).
For according to the definition of similarity, we have that
\(O_{i}^{'} = h*O_{i}\) for all objects in the set, for some
diffeomorphism \(h\). Thus, if \(\left\langle x_{i} \right\rangle\) is
well-adapted to \(\text{\{}O_{i}\text{\}}\) in \(\mathfrak{M}\), then
the coordinate system \(\left\langle y_{i} \right\rangle\) defined by
\(y_{i} = x_{i} \circ h\) is well-adapted to
\(\text{\{}O_{i}^{'}\text{\}}\) in \(\mathfrak{N}\). Conversely, suppose
that some model \(\mathfrak{M}\) of \(T\) has a coordinate system
\(\left\langle x_{i} \right\rangle\) that is globally well-adapted to a
subset of the spacetime structure fields
\(\text{\{}O_{i}\text{\}} \subset S\), and suppose that in any other
model \(\mathfrak{N}\) of \(T\) there also exists a coordinate system
\(\left\langle y_{i} \right\rangle\) that is globally well-adapted to
the corresponding set of spacetime structure fields
\(\text{\{}O_{i}^{'}\text{\}} \subset S^{'}\). Then there exists a
diffeomorphism \(h:M \rightarrow M\), defined by the condition
\(x_{i}(hp) = y_{i}(p)\), such that \(O_{i}^{'} = h*O_{i}\) for all
objects in the set. Thus, the set of object types
\(\text{\{}\mathbf{O}_{\mathbf{i}}\text{\}} \subset \mathbf{S}\ \)remains
similar for \(T\), and so \(T\) possesses immutable spacetime structure.
We have thus proved that a theory \(T\) will possess a preferred class
of ``well-adapted'' coordinate systems that is applicable \emph{across}
its models if and only if it possesses immutable spacetime structure.

The idea of immutable spacetime structure allows us to explicate another
idea associated with GR's distinctive character: that of ``background
independence''. There is wide agreement that GR is distinctive in
lacking certain kinds of structures that serve as a fixed background to
the physical goings-on in all the models. The debate revolves around the
question of how exactly we are to understand this concept of a ``fixed
background''. One suggestion is that background
structures should be identified with non-dynamical fields. But we saw
earlier that, at least under one plausible construal of what it takes
for a field to be ``non-dynamical'', this suggestion may deliver
competing verdicts when applied to different formulations of the same
theory. If we instead identify fixed background structures with
immutable spacetime structure, we avoid this problem. As a bonus, we
also elicit the connection between GR's background independence and its
lack of preferred global coordinate systems applicable across its models, and hence with its
general-covariant formulation.

\section{Conclusion}
To conclude, I have argued that diffeomorphism invariance is not what is
distinctive about GR, for under any definition of the concept according
to which it can be predicated of \emph{theories} rather than
\emph{formulations of theories}, diffeomorphism invariance does not
drive a wedge between GR and other classical spacetime theories.

Nevertheless, the properties of a theory can place constraints on the
possibilities for its formulation. The properties of GR are such that
the theory can only be given a general-covariant formulation, which
makes its diffeomorphism invariance (in the sense of
diff-invariance\textsubscript{1}) manifest. GR must be formulated in
this way because it lacks even a single preferred coordinate system to
which the equations can be referred, and in which they simplify. The
underlying reason for this can be traced back to two distinctive
features of general relativistic spacetimes: (1) the spacetime structure
is mutable \emph{across} models, as expressed mathematically by the
non-existence of any non-empty subset of spacetime structure fields that
remain similar for GR; and (2) the spacetime structure is generically
highly inhomogeneous \emph{within} models, as expressed mathematically
by the (generic) non-existence of a global geodesic congruence of
curves. It is therefore these latter properties which are ultimately
responsible for the distinctive character of general relativity as a classical spacetime theory.

\backmatter

\newpage
\bibliography{citations.bib}

\end{document}